\documentclass[aps,prl,reprint,superscriptaddress]{revtex4-2}

\usepackage{graphicx}
\usepackage{todonotes}
\usepackage{amsmath}
\usepackage{soul,xcolor}
\usepackage{hyperref}
\usepackage[T1]{fontenc}
\usepackage{float}
\usepackage{bm}
\usepackage{amssymb}

\begin{document}

\title{Magnetic Control of Stochastic Spin Selection in a Trapped Polariton Condensate}

\author{Meng Niu}
\email{niumeng@westlake.edu.cn}
\affiliation{Department of Physics, Fudan University, Shanghai 200433, China}
\affiliation{Department of Physics, School of Science, Westlake University, Hangzhou 310014, China}
\author{Igor Chestnov}
\email{igor.chestnov@metalab.ifmo.ru}
\affiliation{School of Physics and Engineering, ITMO University, St. Petersburg 191002, Russia}
\author{Xiaoqing Zhou}
\affiliation{Department of Physics, School of Science, Westlake University, Hangzhou 310014, China}

\author{Pavlos G. Savvidis}
\affiliation{Department of Physics, School of Science, Westlake University, Hangzhou 310014, China}

\begin{abstract}
  Bistability provides a physical basis for memory and switching and its operation depends on the ability to control which of the coexisting states is selected. We demonstrate combined magnetic and optical control of polarization selection in an exciton-polariton condensate confined in an optical trap. Full Stokes polarimetry shows that under linearly polarized nonresonant pumping the condensate randomly acquires right or left circular polarization above threshold. Pump ellipticity biases this selection and changes a bimodal distribution into a unimodal one, while a magnetic field restores equal probabilities by compensating the pump-induced spin imbalance of the reservoir. At higher powers the bimodal distribution vanishes, with a reduction in the degree of polarization that a stochastic model relates to temporal polarization fluctuations. These findings establish magnetic compensation as a means to restore balanced polarization states in a driven condensate and define the regime where both remain accessible.
   
\end{abstract}

\maketitle

\textit{Introduction.}---Bistability, the coexistence of two stable states in a driven nonlinear system, provides the physical basis for optical memory and switching elements \cite{Baas2004,Gippius2007,Cerna2013}. Their operation requires stable coexisting states and control over which one is selected. Exciton polaritons, hybrid quasiparticles formed by the strong coupling of quantum well excitons to microcavity photons \cite{Weisbuch1992,Carusotto2013,Deng2010}, are a natural platform for such elements, since they combine strong nonlinearities and fast dynamics with direct optical access to their state \cite{Sanvitto2016,Amo2010,Ballarini2013}. Above threshold, polaritons condense into a macroscopic quantum state \cite{Kasprzak2006,Deng2002}. Under non-resonant linearly polarized excitation, an optically trapped condensate spontaneously adopts one of two highly circularly polarized states of opposite handedness with equal probability, so that its spin is bistable \cite{Ohadi2015,Redondo2018,Balas2022}. This spin bifurcation can be interpreted as spontaneous breaking of the spin symmetry when spin-dependent interactions destabilize the linearly polarized condensate \cite{Ohadi2015}. Deterministic manipulation of these bistable spin states has been demonstrated through electrical switching at energies below 0.5~fJ \cite{Dreismann2016}, while optical tuning of the coupling between condensates controls spin alignment in pairs  \cite{Ohadi2016PRL} and spin order in chains \cite{Ohadi2017PRL}.

\begin{figure}[!b]
    \centering
    \begin{tikzpicture}[
        label/.style={font=\footnotesize, anchor=south west}
    ]
        \node[anchor=west] (figure 1) at (-0.2,0) {\includegraphics[width=0.47\textwidth]{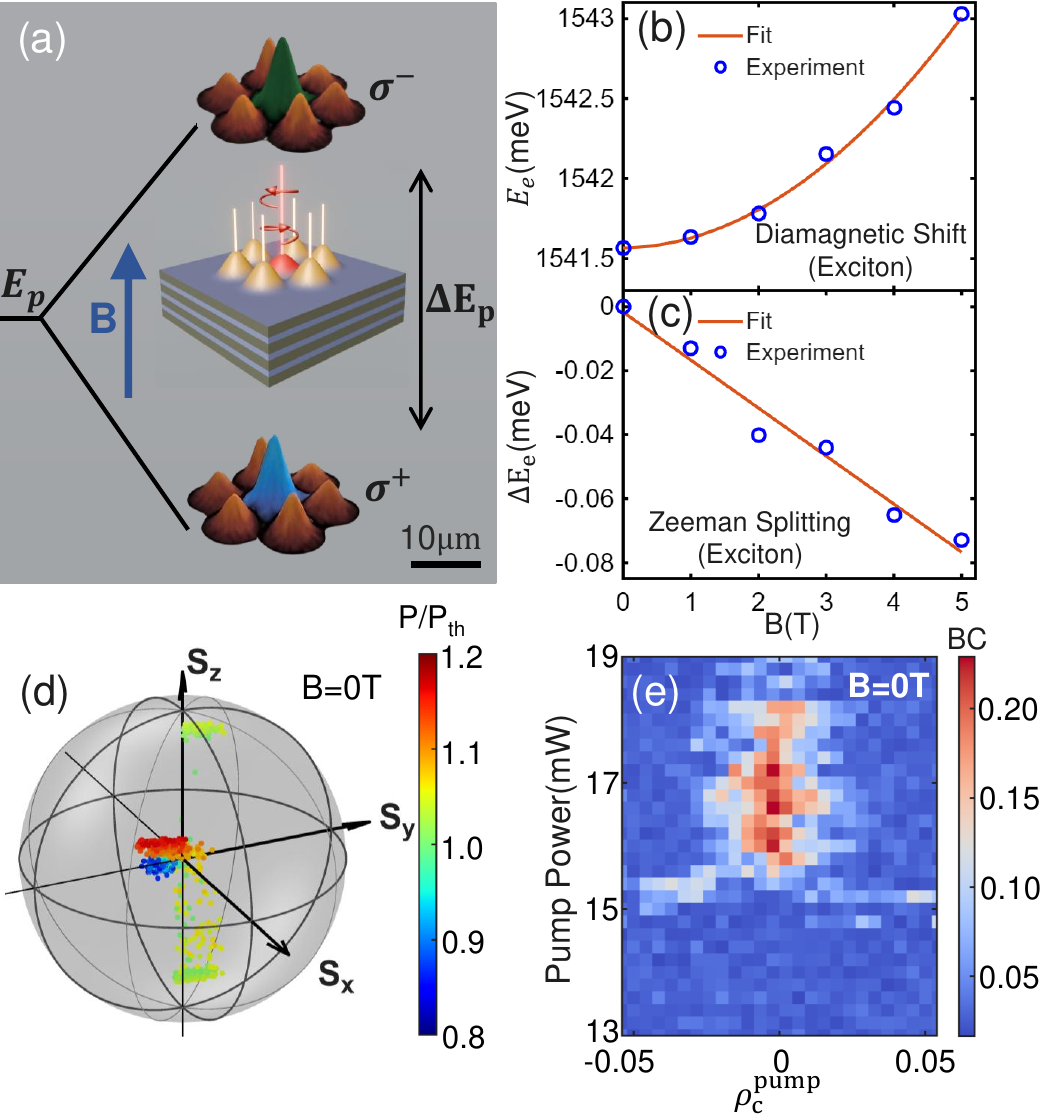}};
    \end{tikzpicture}
    \caption{
    (a) Schematic of the experiment. A nonresonant pump beam is shaped by a spatial light modulator (SLM) into a hexagonal pattern of six Gaussian spots that excites the microcavity placed in an applied magnetic field $B$ (normal to the cavity plane). Exciton polaritons trapped at the center of the hexagon stochastically condense into either $\sigma^{+}$ or $\sigma^{-}$ state, Zeeman split by $\Delta E_p$. 
    (b),(c) Measured and fitted (b) diamagnetic shift and (c) Zeeman splitting of the excitons in an external magnetic field. 
    (d) Condensate polarization at $B = 0$ under linearly polarized pumping, mapping on the Poincar\'e sphere via the Stokes parameters ($S_x, S_y, S_z$) and colored by normalized pump power $P/P_{\mathrm{th}}$. 
    (e) Map of the condensate bistability, quantified by Hartigan's dip test of unimodality \cite{hartigan1985}, versus pump power and pump circular polarization degree $\rho_c^{\rm pump}$.}
    \label{fig:1}
\end{figure}

Controlling the statistics of spontaneous spin selection requires a perturbation that breaks the equivalence of the two spin states. A magnetic field normal to the cavity plane lifts spin degeneracy by introducing a Zeeman splitting between the circularly polarized polariton components $\sigma_+$ and $\sigma_-$ \cite{Chinh2024,Bochin2024,PhysRevB.107.125303,NguyenTruong2023,Lundt2019,RahimiIman2011,PhysRevLett.106.257401,Krol2018}. In terms of condensate pseudospin, the Zeeman term acts as a longitudinal field analogous to the local field in an Ising model \cite{Kalinin2020}. Although magnetic manipulation of polariton spins has been regarded as requiring strong fields \cite{Dreismann2016}, near a spin bifurcation, even a weak magnetic field is predicted to substantially bias spontaneous spin selection \cite{Bochin2024}. However, the Zeeman splitting in optically trapped condensates can vanish or even reverse sign depending on the pump power \cite{Sawicki2024}. In addition, an elliptically polarized pump can also bias spin selection toward the condensate state of the same handedness \cite{Ohadi2015}. This raises two questions. First, how does the interplay between the magnetic field and pump ellipticity affect spin selection, and can these effects compensate for each other to restore equal probabilities? Second, over what range of fields and pump powers do both spin states remain accessible?

Here we address these questions for an optically trapped polariton condensate in a GaAs based microcavity. Using full Stokes polarimetry under quasi continuous pumping, we map the polarization statistics of the condensate as a function of pump polarization, pump power, and magnetic field. Fields below 1~T shift the point of equal probability to a finite pump ellipticity whose sign and magnitude depend on the sign and strength of the field, so that an elliptically polarized pump restores balanced spin selection, while the bistable range narrows with increasing field and eventually vanishes. A stochastic two-mode model with a spin dependent reservoir reproduces these maps, attributes the compensation to the effective field of the spin polarized reservoir opposing the Zeeman splitting, and relates the loss of the total degree of polarization at high power to temporal polarization fluctuations.


\textit{Results.}---Experiments were carried out on a $5\lambda/2$ microcavity containing 12 Al$_{0.3}$Ga$_{0.7}$As/GaAs quantum wells, held at $\approx 9$~K. The cavity was excited nonresonantly at 749~nm with 1~ms quasi cw pulses at a repetition rate of 10~Hz to suppress heating (see Supplemental Material). A spatial light modulator shaped the pump into six Gaussian spots on a hexagon of 10~$\mu$m radius [Fig.~\ref{fig:1}(a)], and the resulting reservoir forms an in plane trap that confines the condensate at its center. The emission was analyzed with a full Stokes polarimeter that records all four Stokes parameters in every shot.
\begin{figure}[t]
    \centering
    \includegraphics[width=0.48\textwidth]{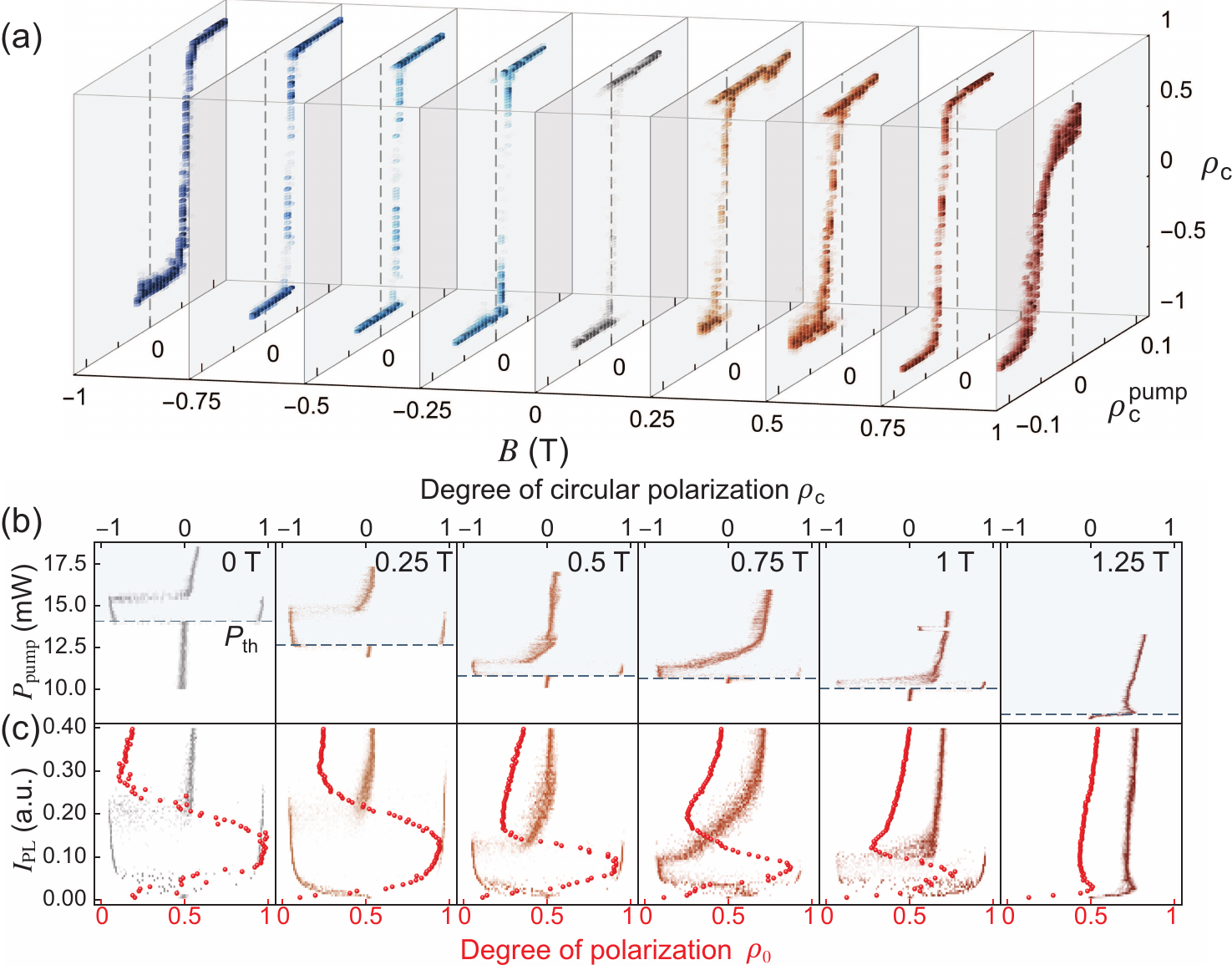}
    \caption{
    (a) Probability distribution of the condensate circular polarization degree $\rho_{\rm c}$ versus pump circular polarization degree $\rho_{\rm c}^{\rm pump}$ around linearly polarized pumping ($\rho_{\rm c}^{\rm pump} = 0$), measured at $P_{\rm pump}\approx1.1P_{\rm th}$ for magnetic fields from $-1$ to $1$~T. For each value of $\rho_{\rm c}^{\rm pump}$, 100 single-shot realizations were recorded, with the color intensity giving the probability of observing each value of $\rho_{\rm c}$.
    (b) Probability distribution of $\rho_{\rm c}$ versus pump power around the condensation threshold, measured at the balance point (the value of $\rho_{\rm c}^{\rm pump}$ at which $\sigma^+$ and $\sigma^-$ occur with equal probability), for magnetic fields from $0$ to $1.25$~T (left to right columns).
    (c) Data of (b) rearranged versus the PL intensity instead of the pump power. Red points (bottom axis) show the total degree of polarization $\rho_0$.}
    \label{fig:2}
\end{figure}

\begin{figure*}[t]
    \centering
    \includegraphics[width=1\textwidth]{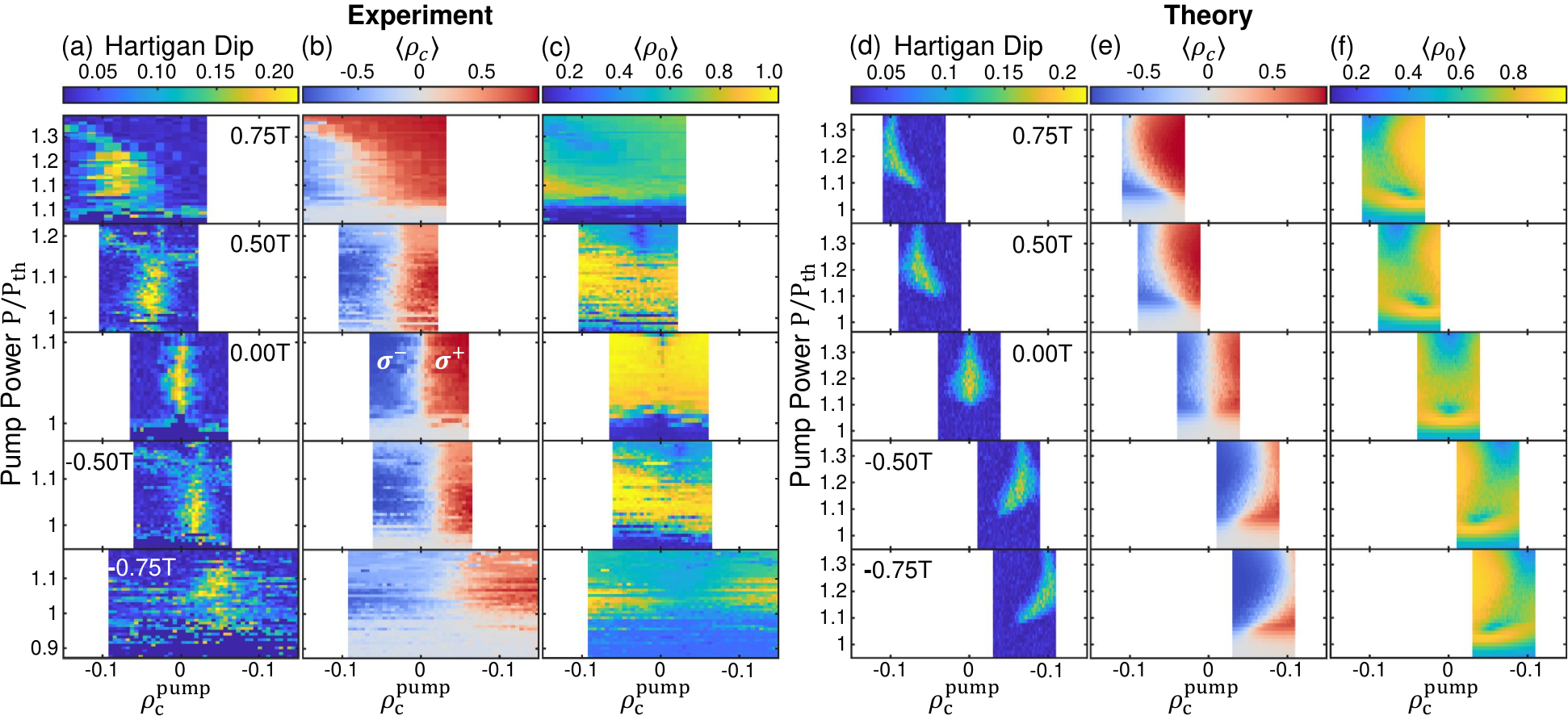}
\caption{
    Bistability maps of the trapped condensate versus pump polarization and pump power.
    (a) Hartigan dip statistic of the $\rho_{\rm c}$ distribution.
    (b) Ensemble averaged circular polarization $\langle\rho_{\rm c}\rangle$.
    (c) Ensemble averaged total degree of polarization $\langle\rho_0\rangle$, versus $\rho_{\rm c}^{\rm pump}$ and pump power for magnetic fields between $-0.75$ and $0.75$~T. Larger dip values indicate a more pronounced bimodal distribution. 
    In (b), blue and red regions correspond to condensates with $\sigma^{-}$ and $\sigma^{+}$ polarization, respectively. The pump power is normalized to the threshold $P_{\mathrm{th}}$ measured at the balance point, where $\sigma^{+}$ and $\sigma^{-}$ occur with equal probability.
    (d-f) Numerically simulated polarization statistics of the trapped condensate, which match the experimental data in (a-c).
    }
    \label{fig:3}
\end{figure*}

Fits of a coupled-oscillator model to the lower and upper polariton spectra measured well below threshold yield the energies $E_{\mathrm{ex},+}$ and $E_{\mathrm{ex},-}$ of the bright exciton spin states associated with $\sigma^{+}$ and $\sigma^{-}$ emission, respectively (see Supplemental Material).  In a magnetic field $B$, the mean exciton energy follows $E_{\mathrm{ex}} = E_{0} + \alpha B^{2}$ with $\alpha = 0.0564~\mathrm{meV/T^{2}}$ and the exciton Zeeman splitting is $\Delta E = E_{\mathrm{ex},+} - E_{\mathrm{ex},-} = g\mu_B B$ with $g = -0.274$ [Figs.~\ref{fig:1}(b,c)]. By lifting the exciton spin degeneracy, the magnetic field thus couples directly to the condensate spin.

Using the polarimetry setup, we characterized the polarization state of the condensate by the Stokes parameters, defined as $S_0=I_{\sigma^+}+I_{\sigma^-}, S_x=(2I_H-S_0)/S_0,S_y=(2I_D-S_0)/S_0,S_z=(I_{\sigma^+}-I_{\sigma^-})/S_0$ and mapped them onto the Poincar\'e sphere [Fig.~\ref{fig:1}(d)]. Here H and D refer to the horizontal and diagonal linear polarizations, respectively. At zero magnetic field and under linearly polarized pumping, the PL below threshold is unpolarized (blue points near the origin of the sphere), since no condensate has formed. Just above threshold ($P \approx 1.05\,P_{\mathrm{th}}$), the condensate becomes strongly circularly polarized and randomly selects one of the two poles of the sphere from shot to shot. We refer to this behavior as the spin bistability of the exciton polariton condensate.

To quantify the bistability, we apply Hartigan's dip test of unimodality~\cite{hartigan1985} to the measured distribution of the circular polarization degree. The test discriminates between unimodal and multimodal distributions, and the resulting dip statistic serves as a measure of the strength of bimodality. As the map in Fig.~\ref{fig:1}(e) shows, bistability occurs only within a narrow window of pump polarization around $\rho_c^{\rm pump}=0$ and of pump power near threshold, between 15 and 18~mW.

\textit{Polariton polarization bistability}---Since the Zeeman splitting lifts the exciton spin degeneracy, spin relaxation toward the lower Zeeman level produces a small imbalance between the $\sigma^{+}$ and $\sigma^{-}$ populations of incoherent high-momentum excitons (the reservoir) even under linearly polarized pumping. In addition, the polariton Zeeman splitting biases the dynamics during condensate formation, so that small polarization fluctuations lead to one spin state more often than to the other \cite{Bochin2024}.  To study how the field affects spin selection in the condensate, we sweep the pump power and pump polarization at a series of magnetic fields while recording the full Stokes parameters of the PL.

Fig.~\ref{fig:2}(a) shows the circular polarization degree of the condensate, $\rho_{\rm c}$, as a function of the pump circular polarization degree $\rho^{\rm pump}_{\rm c}$. For each value of $\rho^{\rm pump}_{\rm c}$, we acquired 100 single-shot measurements of $\rho_{\rm c}$, and the color scale gives the probability of observing each value. In a symmetric case realized at $B = 0$ and under linearly polarized pumping ($\rho^{\rm pump}_{\rm c} = 0$), the $\sigma^{+}$ and $\sigma^{-}$ reservoir populations are equal, and the condensate settles into either the $\sigma^{+}$ or the $\sigma^{-}$ state with equal probability. In the middle panel of Fig.~\ref{fig:2}(a) this appears as two branches near $\rho_{\rm c} = \pm 1$ that coexist around $\rho^{\rm pump}_{\rm c} = 0$.

A finite magnetic field breaks this symmetry and shifts the balance point, defined as the value of $\rho^{\rm pump}_{\rm c}$ at which both circular polarizations occur with equal probability. The direction and magnitude of the shift depend on the sign and strength of the field, showing that the field acts as a sensitive bias on the spin selection. Since $g < 0$, a positive field lowers the $\sigma^{+}$ exciton energy and favors the $\sigma^{+}$ reservoir population, so a pump with a small negative $\rho^{\rm pump}_{\rm c}$ is needed to restore the balance.

Figure~\ref{fig:2}(b) shows the distribution of $\rho_{\rm c}$ as a function of pump power near threshold. The magnetic field reduces the condensation threshold and narrows the bistable range, which vanishes completely at $B = 1.25$~T, where the condensate locks into a single spin state. 

Finally, Fig.~\ref{fig:2}(c) presents the data of Fig.~\ref{fig:2}(b) sorted by PL intensity, together with the total degree of polarization $\rho_0$ (red points). Since the PL intensity rises steeply at threshold, this representation expands the region just above threshold and resolves how the polarization builds up. The onset of condensation is accompanied by a continuous increase of $\rho_0$ from zero, which reaches its maximum once the condensate is fully circularly polarized and the bistable state is established. At higher condensate populations, the spin no longer remains at a fixed point but follows complex precessional trajectories on the Poincar\'e sphere, including circulation along limit cycles and noise induced switching between the two hemispheres (see the theoretical analysis below). Averaged over the measurement window, these dynamics reduce $\rho_0$ and, at the same time, suppress the bistable response.

The spin bistability of the condensate is governed by the pump polarization, the pump power and the magnetic field, and we now examine how these three parameters act together. We scan the pump polarization and pump power at several magnetic fields and map the bimodality, average circular $\langle\rho_{\rm c}\rangle$ and total polarization degree $\langle\rho_0\rangle$ as two dimensional diagrams in Figs.~\ref{fig:3}(a-c).

The bimodal region is confined to a narrow window of $\rho^{\rm pump}_{\rm c}$ around the balance point [Fig.~\ref{fig:3}(a)]. Bistability therefore survives only while the $\sigma^{+}$ and $\sigma^{-}$ reservoir populations remain nearly equal, and even a small imbalance drives the condensate deterministically into one of the two spin states. As the field increases, this window shifts along the $\rho^{\rm pump}_{\rm c}$ axis, following the compensation of the Zeeman splitting by the pump induced reservoir spin imbalance.

The maps of $\langle\rho_{\rm c}\rangle$ and $\langle\rho_0\rangle$ [Figs.~\ref{fig:3}(b,c)] support this picture. The white region in Fig.~\ref{fig:3}(b), where $\langle\rho_{\rm c}\rangle$ vanishes, coincides with the region of high bimodality in Fig.~\ref{fig:3}(a) and marks the parameters for which $\sigma^{+}$ and $\sigma^{-}$ occur with equal probability. Near threshold this region coexists with large $\langle\rho_0\rangle$, since each shot remains strongly polarized. At higher pump power $\langle\rho_0\rangle$ decreases and the bimodality eventually disappears. Here the dynamics of the condensate pseudospin are dominated by precession about an effective field that combines the Zeeman splitting with interaction and reservoir contributions, so its strength grows with both the applied field and the condensate population. Temporal averaging over changes in the polarization direction within a single shot leads to a reduction in $\langle\rho_0\rangle$. If fluctuations arising from the condensate's coupling to its environment cause the circular polarization to change sign within a shot, time averaging makes the measured distribution unimodal.


\textit{Model.}---To describe the polarization statistics, we model two trapped modes of opposite circular polarization, $\psi_{\pm}$, coupled to spin dependent incoherent reservoirs $n_{\pm}$ by stochastic Langevin equations~\cite{Ohadi2015,Ryzhov2020,Balas2022}. The model includes the polariton Zeeman splitting $\hbar\Delta$, same and opposite spin interactions within the condensate and with the reservoir, spin dependent stimulated scattering from the reservoir, the energy and loss splitting $\varepsilon + i\sigma$ between the linearly polarized modes, and a field-induced asymmetry of the reservoir spin relaxation, $\xi = -\tanh[g\mu_{B}B/(2k_{B}T)]$. The full equations, noise terms and parameters are given in the Supplemental Material. In terms of the pseudospin $\mathbf{s}$, whose normalized time average gives the measured Stokes vector $\mathbf{S}$, the conservative dynamics consist of precession about the effective field $\bm{\Omega} = (\varepsilon, 0, \Omega_z)$ with
\begin{equation}
\Omega_z = \Delta + 2(\alpha_1 - \alpha_2)s_z + (\alpha_{1r} - \alpha_{2r})m,
\label{eq:model-effective-splitting}
\end{equation}
where $\alpha_{1,2}$ and $\alpha_{1r,2r}$ are the condensate and reservoir interaction constants and $m = n_{+} - n_{-}$ is the reservoir spin imbalance. The last term is controlled by the pump polarization, since $m \propto \rho_{\rm c}^{\rm pump}$, and can oppose the Zeeman splitting. Coexistence of the two circular polarizations should therefore persist at finite field, with the bistable region shifted towards the pump polarization that compensates the Zeeman term.

We simulate the stochastic dynamics over 200~ns and compute for each realization the time averaged circular and total degrees of polarization, $\rho_{\rm c} = \overline{s_z}/\overline{s_0}$ and $\rho_0 = |\overline{\mathbf{s}}|/\overline{s_0}$, and from many realizations the dip statistic, $\langle\rho_{\rm c}\rangle$ and $\langle\rho_0\rangle$ (see Supplemental Material). Here $s_0$ is the total condensate population and the overbar denotes time averaging. The calculated maps [Figs.~\ref{fig:3}(d-f)] reproduce the main features of experimental data in Figs.~\ref{fig:3}(a-c). The bimodal region is confined to a narrow window of pump polarization and shifts along the $\rho_{\rm c}^{\rm pump}$ axis with the field in the direction predicted by the compensation mechanism, while near the balance point $\langle\rho_{\rm c}\rangle$ vanishes and $\langle\rho_0\rangle$ decreases at high pump power.

\begin{figure}
    \centering
    \includegraphics[width=0.9\linewidth]{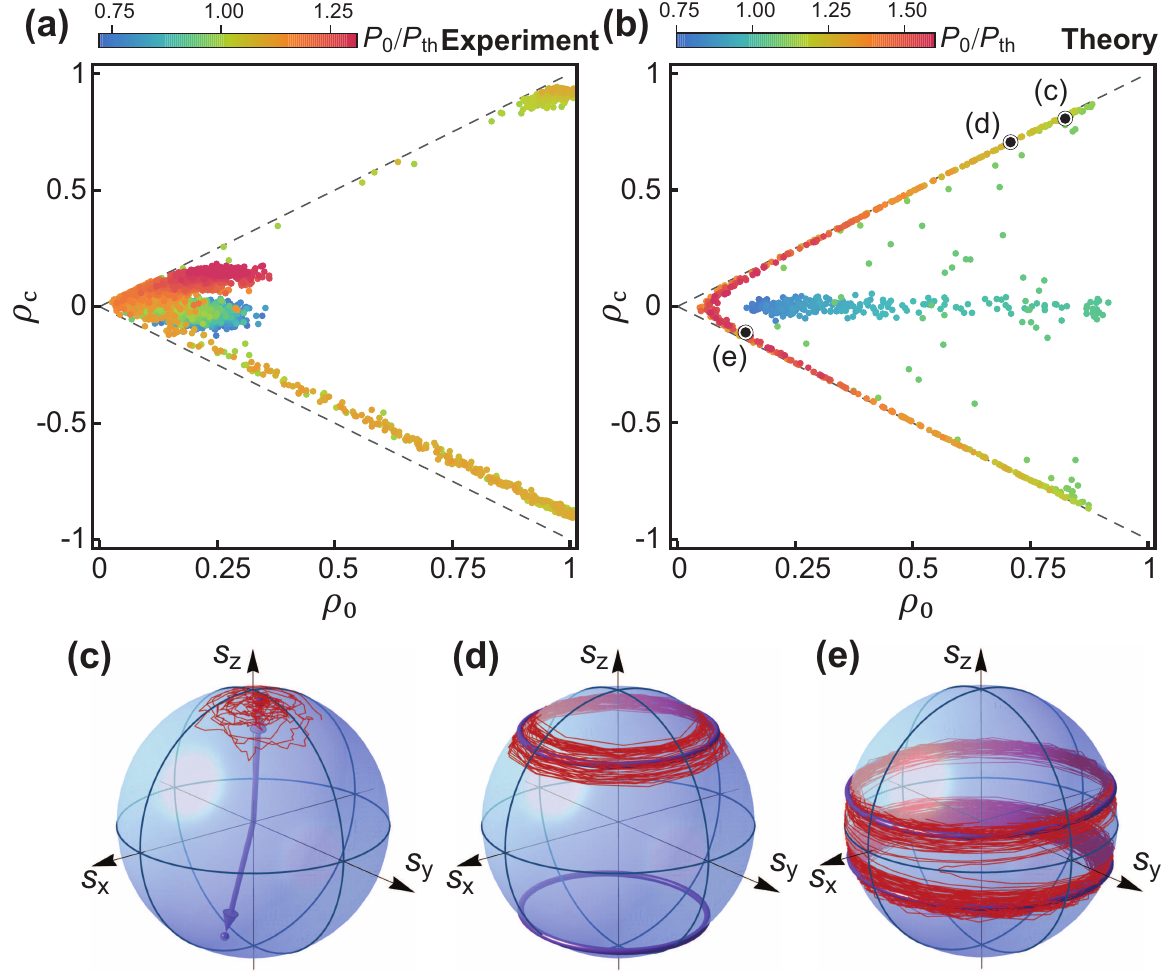}
    \caption{Polarization dynamics of the trapped condensate. (a,b) Experimental and calculated circular polarization $\rho_c$ versus total polarization $\rho_0$, respectively. Colors indicate the pump power $P_0$, normalized to the condensation threshold $P_{\mathrm{th}}$. Dashed lines correspond to $\rho_c=\pm\rho_0$.  (c--e) Trajectories of the normalized pseudospin on the Poincar\'e sphere at $P_0/P_{\mathrm{th}}=1.10,\ 1.24,\ 1.52$, respectively. Purple arrows in (c) indicate the symmetry-broken stationary states; thick purple curves in (d,e) show limit cycle solutions. Thin red curves show stochastic trajectories near a stationary state over time intervals of 6, 6, and 20 ns for panels (c--e), respectively.}
    \label{fig:StocksDyn}
\end{figure}

Fig.~\ref{fig:StocksDyn} shows the polarization dynamics underlying this depolarization at $B = 0$ under linearly polarized pumping. Close to threshold, the deterministic evolution converges to one of two fixed points with opposite circular polarization [Fig.~\ref{fig:StocksDyn}(c)], while the stochastic trajectory remains close to the fixed point. At higher power, the trajectories approach one of two limit cycles in opposite hemispheres of the Poincar\'e sphere [Fig.~\ref{fig:StocksDyn}(b)], so the polarization becomes time-dependent while keeping the sign of its circular component. The instantaneous condensate remains fully polarized, but averaging a Stokes vector whose direction changes within the integration window reduces $\rho_0$ of each individual shot. The periodic trajectories are almost parallel to the equatorial plane, so the linear polarization degree vanishes under time averaging while $\rho_c \approx \rho_0$. This agrees with the data from the pump power sweep [Fig.~\ref{fig:2}(b,c)], plotted in the $(\rho_c,\rho_0)$ plane, -- see Fig.~\ref{fig:StocksDyn}(a). Our simulations reproduce this behavior reasonably well. Sweeping pump powers reveals a gradual decrease of time-averaged polarization degree, as shown in Fig.~\ref{fig:StocksDyn}(b). Ones the cycles approach the equator, noise-induced switching between the hemispheres sets in [Fig.~\ref{fig:StocksDyn}(e)], reducing $\rho_0$ to near-zero values.. This depolarization is distinct from the cancellation of $\langle\rho_{\rm c}\rangle$ upon ensemble averaging over two stationary states. In addition, it manifests the loss of bimodality and the decrease of $\langle\rho_0\rangle$ observed at high power in Fig.~\ref{fig:3}.

In summary, we have shown that the stochastic spin selection of a trapped polariton condensate can be controlled by combining a weak magnetic field with the pump polarization. Fields below 1~T shift the point of equal spin probability to a finite pump ellipticity, at which the reservoir spin imbalance created by the pump compensates the Zeeman splitting and restores balanced selection. At high pump power, precession of the condensate pseudospin depolarizes individual shots and suppresses the bistability. These results define the parameter window in which both spin states remain accessible, a prerequisite for using polariton condensates as stochastic spin elements in optical memories and Ising type simulators~\cite{Kalinin2020}.

\begin{acknowledgments}
\textit{Acknowledgments.}---P.G.S. acknowledges support from Quantum Science and Technology-National Science and Technology Major Project (2023ZD0300300), the Innovation Resource Allocation of Westlake University-Zhejiang Provincial Natural Science Foundation of China under Grant No. XHD24A2401, and the Westlake University project No. 102510036022301.\\
\end{acknowledgments}

    %
    %
    %

\bibliographystyle{apsrev4-2}
\bibliography{bibliography}
\end{document}